\documentclass[12pt]{iopart}
\pdfoutput=1
\usepackage{subfigure}
\usepackage{iopams}  
\usepackage{float}
\usepackage{url}

\usepackage{graphicx}
\usepackage{bm}
\usepackage[caption=false]{subfig}
\usepackage{color,soul}

\newcommand{\red}[1]{\color{black} #1}

\begin{document}
\title{Coarse-graining in micromagnetic simulations 
of dynamic hysteresis loops} 

\author{R Behbahani$^{1, 2}$, M L Plumer$^1$ and I Saika-Voivod$^{1, 2}$}
\address{1 Department of Physics and Physical Oceanography, Memorial University of Newfoundland, Canada}

 \address{2 Department of Applied Mathematics, University of Western Ontario, London, Ontario, Canada, N6A 3K7
}
\ead{saika@mun.ca}


%
%
%
%


\begin{abstract}
We use micromagnetic simulations based on the stochastic Landau-Lifshitz-Gilbert equation to calculate dynamic magnetic hysteresis loops at finite temperature that are invariant with simulation cell size. As a test case, we simulate a magnetite nanorod, the building block of magnetic nanoparticles that have been employed in preclinical studies of hyperthermia. With the goal to effectively simulate loops for large iron-oxide-based systems at relatively slow sweep rates on the order of 1 Oe/ns or less, 
we modify and employ a previously derived renormalization group approach for coarse-graining (Grinstein and Koch, Phys.~Rev.~Lett. 20, 207201, 2003).  The scaling algorithm is shown to produce nearly identical loops over several decades in the model cell volume. We also demonstrate sweep-rate scaling involving the Gilbert damping parameter that allows orders of magnitude speed-up of the loop calculations. 
\end{abstract}




%
%


The fundamental premise of micromagnetics is that the physics of interest can be modeled by a macrospin representing a collection of atomic spins within a small finite volume, or cell.
The approximation that all spins within a cell point in the same direction is valid at temperature $T=0$, so long as cells remain smaller than the exchange length~\cite{abo2013definition}.  A limiting factor for micromagnetic computer simulations is the number of cells used to model the system; using larger cells is computationally advantageous.


At finite $T$, a few schemes have been proposed to account for how parameters used for modelling the magnetic properties of the material must vary with cell size in order to keep system properties invariant with cell size.  For example, Kirschner et al.~\cite{kirschner2006relaxation, kirschner2005cell} suggested an approximate scaling of saturation magnetization $M_s$ based on the average magnetization of blocks of spins in atomistic Monte Carlo simulations, and subsequently scaling the exchange and uniaxial anisotropy constants $A$ and $K$ to preserve the exchange length and anisotropy field.  Feng and Visscher~\cite{CoarseGrainingFengVisscher} proposed that the damping parameter $\alpha$,
which models the dynamics of magnetic energy loss~\cite{gilbert2004phenomenological}, should scale with cell size, arguing that using larger cells is analogous to having more degrees of freedom for energy absorption; see also~\cite{Wang20112676} for efforts related to $\alpha$.  The renormalization group (RG) approach of Grinstein and Koch~\cite{grinstein2003coarse}, based on mapping a Fourier space analysis of the non-linear sigma model to ferromagnets in order to scale $A$, $K$, field $H$ and magnetization $M$, has garnered significant attention.  However, to the best of our knowledge, no scaling theory has been applied to the calculation of magnetization-field (MH) hysteresis loops~\cite{Westmoreland2018266}, which are the foundation of experimental characterization of magnetic systems.
 
In this Letter, we modify and employ the approach proposed by Grinstein and Koch~\cite{grinstein2003coarse} to the test case of calculating MH loops for magnetite nanorods at sweep rates relevant to magnetic hyperthermia, allowing us to make estimates of specific loss power that would otherwise be computationally impractical.
 

    
The magnetite nanorods we simulate are the building-blocks of the nanoparticles that were shown by Dennis {\it et al} to successfully treat cancerous tumours in mice via hyperthermia~\cite{dennis2009nearly}. It is reasonable to choose the smallest micromagnetic cell to be the cubic unit cell, which is of length $a_0=0.839$~nm and contains 24 magnetic Fe ions. 
We set the exchange stiffness constant to $A_0=0.98\times10^{-11}$~J/m, which for cell length $a_0$ yields an effective exchange constant between neighbouring cells of $J_{\rm eff}=a_0 A_0=8.222\times10^{-21}$~J, which in turn yields a bulk critical temperature of $T_c=1.44 J_{\rm eff}/k_B = 858$~K for the bulk 3D-Heisenberg-model version of our system. This value of $A_0$ is close to what can be theoretically determined by considering the atomic-level exchange interactions across the faces of neighbouring unit cells~\cite{victora2003effects}, and is in reasonable agreement with experimental values~\cite{heider1988note, kouvel1956specific, moskowitz1987theoretical, glasser1963spin, srivastava1979exchange, srivastava1987spin, uhl1995first}.  The nanorod dimensions are approximately 6.7 nm $\times$ 20 nm $\times$ 47 nm ($8a_0 \times 24a_0 \times 56 a_0$), with its length along the $z$-axis.
 We set  {$M_s=480$~kA/m}~\cite{heider1988note, dutz2013magnetic, usov2013properties}, the bulk value for magnetite.
{\red We do not consider magnetostatic interactions explicitly, but rather implicitly through an effective uniaxial anisotropy.  For the purposes of this study, we choose a strength of $K_0 = 10$~kJ/m$^3$, 
which is consistent with other studies of iron oxide nanoparticles~\cite{usov2017interaction, plumer2010micromagnetic}, and 
for which a more precise estimate can be obtained by considering the nanorod's demagnetization tensor~\cite{usov2013properties,cullity2011introduction,Newell1993,Aharoni1998,fukushima1998volume,usov2012magnetic}, maghemite content~\cite{dennis2009nearly}, and the effect of neighbouring nanorods within a nanoparticle.  }
{\red We omit cubic crystalline anisotropy as it has negligible effects on the hysteresis loops of magnetite nanoparticles with even modest aspect ratios, as discussed in Refs.~\cite{ usov2013properties,usov2012magnetic} (we have also verified that adding cubic anisotropy of strength 10~kJ/m$^3$ has no impact on the loops presented here).}
Anisotropy is set along the $z$-axis with a 5$^{\circ}$ dispersion   
to mimic lattice disorder~\cite{plumer2010micromagnetic}.
 For convenience we set $\alpha=0.1$, a choice consistent with previous studies~\cite{plumer2010micromagnetic, usov2010low} and with magnetite thin films~\cite{serrano2011thickness}.

While hysteretic heating is at the heart of magnetic nanoparticle hyperthermia, preventing eddy current heating of healthy tissue limits the frequency $f$ and amplitude $H_{\rm max}$ of the external field such that the sweep rate ${\rm SR}=4 H_{\rm max} f$ is less than a target value of $0.25$~Oe/ns~\cite{hergt2007magnetic, dutz2013magnetic}.  For our simulation, we set $H_{\rm max}=500$~Oe, which for the target SR implies a target value of $f=125$~kHz, a value large enough to restrict unwanted Brownian relaxation~\cite{dutz2013magnetic}.

\begin{figure}
    \centering
    \includegraphics[width=0.8\columnwidth ]{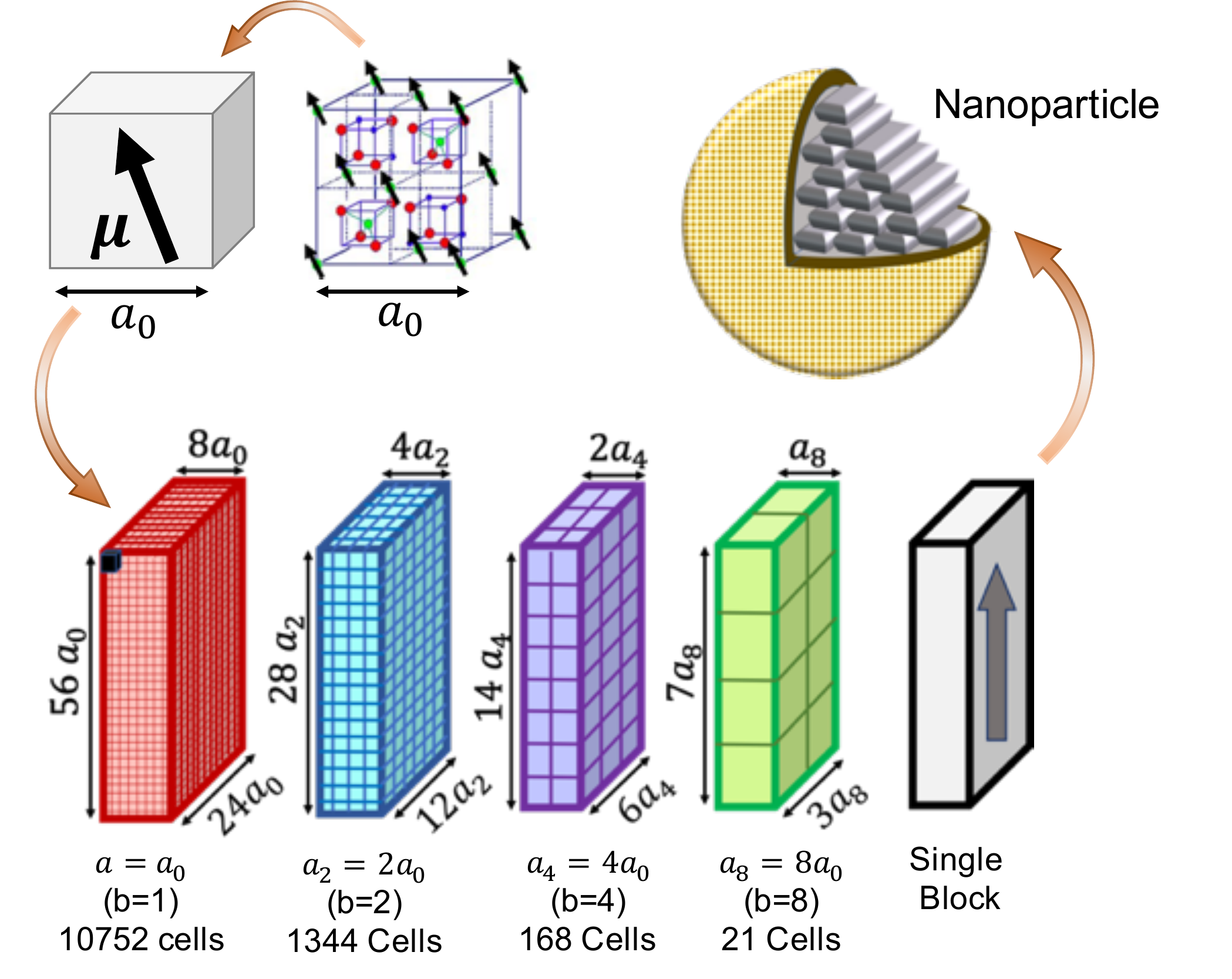}
    \caption{
    Coarse-grained modelling of a magnetite nanorod.  The smallest micromagnetic cell models the atomic spins within a cubic unit cell of length $a_0=0.839$~nm with a single magnetic moment. Our goal is to model the system using a smaller number of larger cells (of length $a_b=b\,a_0$ for $b>1$) with appropriately scaled parameters.
    The number of cells drawn and their sizes are only approximate. Illustrative spins for half of the tetrahedral Fe$^{3+}$ sites (FCC sites) are drawn over a spinel unit cell taken from Ref.~\cite{invspinel}.
    }\label{fig:micromagnetics}
\end{figure}

    
To model the dynamics of the magnetization of a cell $\mathbf{M}$ of fixed magnitude $M_s$, we solve the Landau-Lifshitz-Gilbert (LLG) equation~\cite{cullity2011introduction, gilbert2004phenomenological, brown1963thermal}, 
\begin{equation}
    \frac{d\mathbf{M}}{dt}=-\gamma_1 \mathbf{M} \times \mathbf{H}_{\mathrm{eff}} - \frac{\alpha \gamma_1}{\mathrm{M}_s} \mathbf{M}\times (\mathbf{M}\times \mathbf{H}_{\mathrm{eff}})
\end{equation}
where $t$ is time, $\gamma_1=\mu_0\gamma_e/(1+\alpha^2)$, $\gamma_e=1.76\times10^{11}$ rad/(s.T) is the gyromagnetic ratio for an electron, $\mu_0$ is the vacuum permeability,
and $\mathbf{H}_{\mathrm{eff}}$ is due to the combination of an external field, uniaxial anisotropy, exchange interactions and a thermal field.  
We perform our simulations using OOMMF (Object Oriented Micromagnetic Framework) software~\cite{OOMMF}.
In particular, we include the Theta Evolve module~\cite{theta_evolve} used for simulations at finite $T$ via a stochastic thermal field~\cite{brown1963thermal}.


We simulate the rod using cubic cells of length $b a_0$, with $b$ taking on values 1, 2, 4 and 8. See Fig.~\ref{fig:micromagnetics}. For $b=1$, 10752 cells make up the rod.  For $b=2$, there are $10752/2^3 = 1344$ cells. 
{\red  The  volume  of  the  rod  is fixed for all simulations at $10752 a_0^3 \approx (22a_0)^3$.  Additionally, we  simulate  the  rod  as  a  single  cell  –  a single  rectangular  prism,  or  block. While there is some ambiguity in assigning a single length scale to represent a rectangular prism, we choose $b=22$ from the geometrical mean, i.e., the side length of the cube of the same volume as the rod.}

The goal of coarse-graining is to determine $A(b)$ and $K(b)$, i.e., how the exchange and anisotropy parameters should change with $b$ to keep system properties invariant with $b$.
{\red The $b=22$ case is a practical limit where all the atomic spins are represented by a single macrospin, where exchange interactions are no longer required in the simulations, and which provides for an interesting test of a coarse-graining procedure in predicting $K(b)$.}
In calculating hysteresis loops for a system with cell length $ba_0$, we apply an external field along the $z$ axis of $H(b) = H_{\rm max} \sin{(2 \pi f t)}$, 
and report the $z$-component of the average (over cells) magnetization unit vector $m_H = \bar{M}_z/M_s$, averaged over 88 to 100 independent simulations for $b>1$. For
 $b=1$ we use 250 simulations.

\begin{figure*}
\includegraphics[width = 0.33\columnwidth]{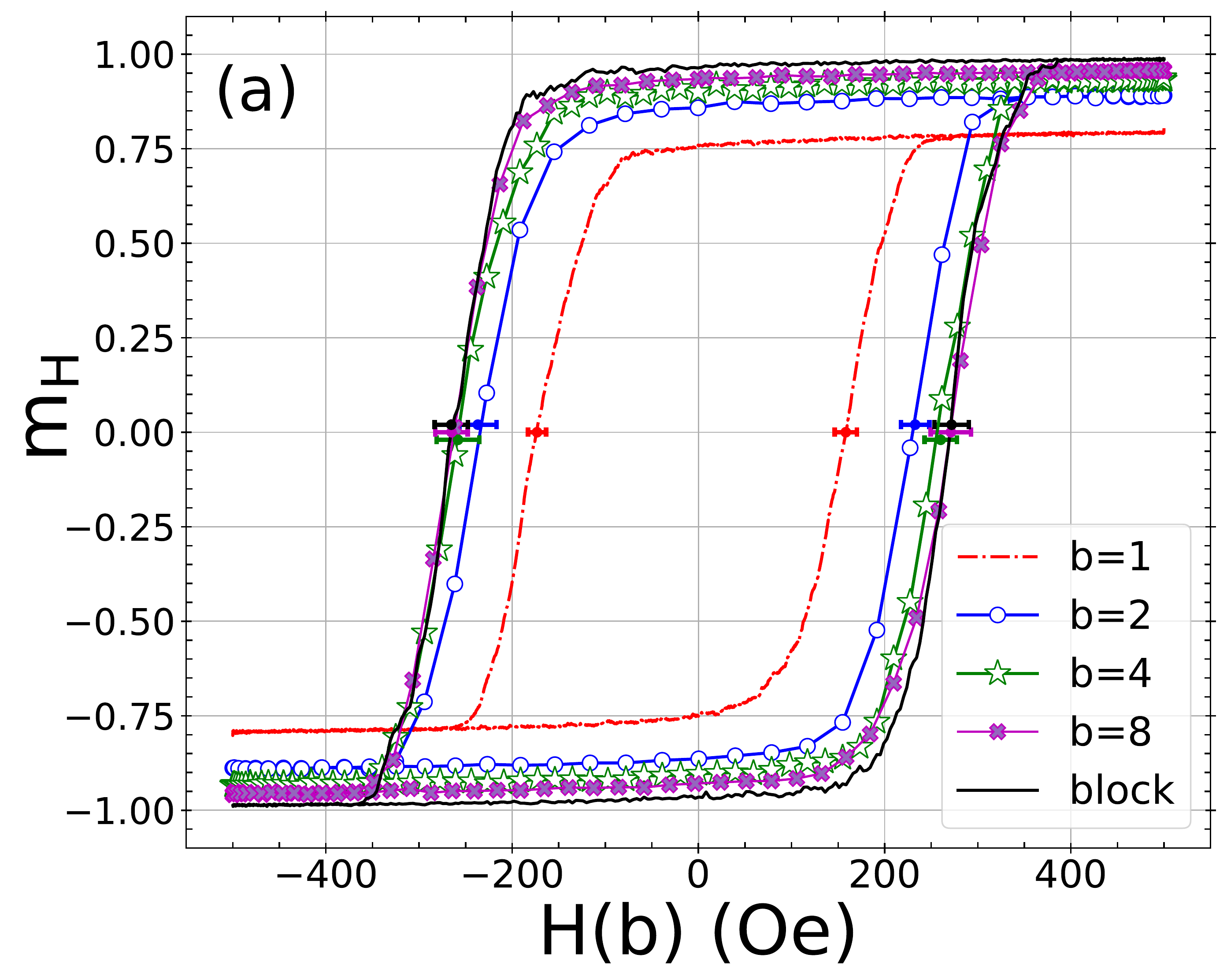}
\includegraphics[width = 0.33\columnwidth]{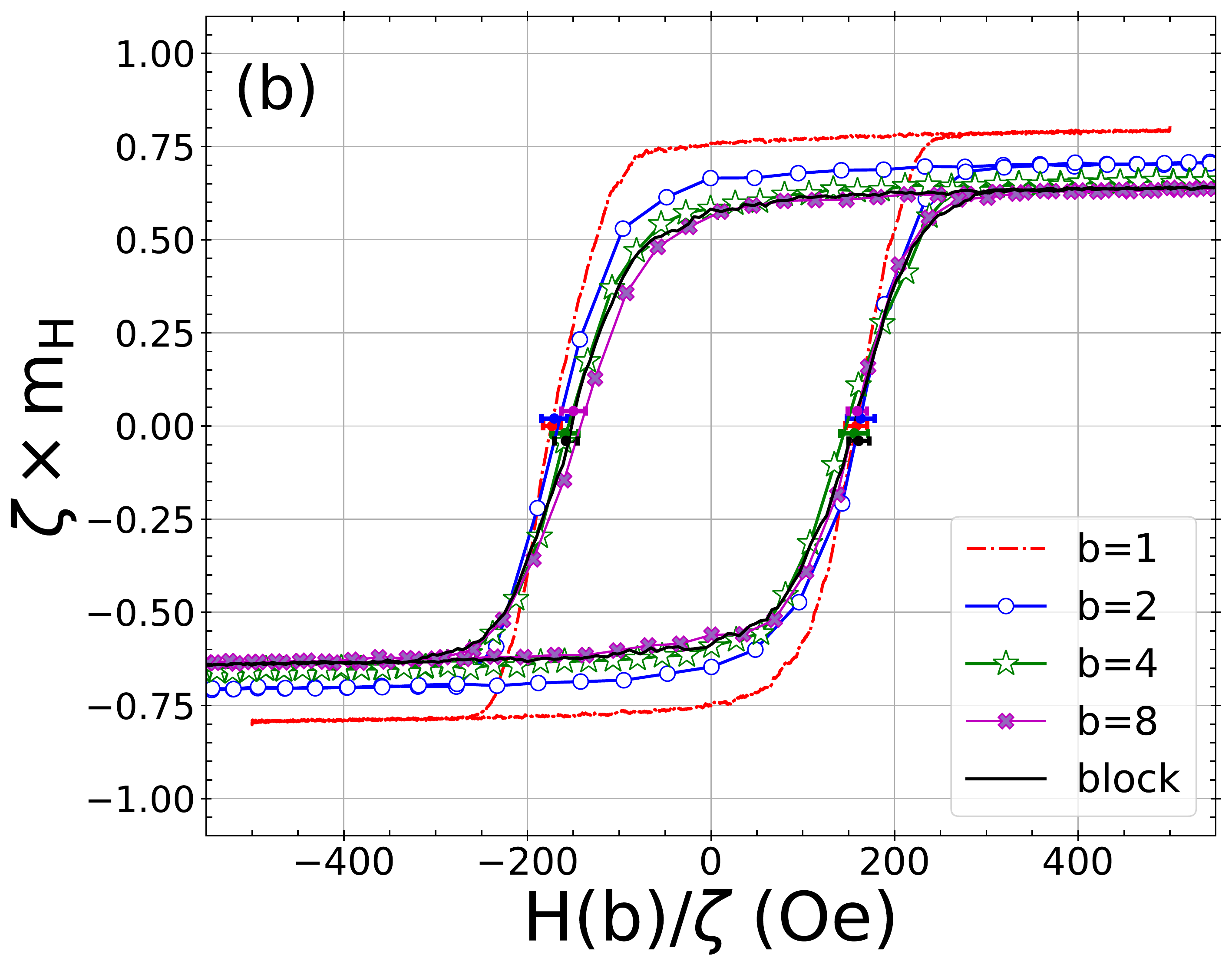}
\includegraphics[width = 0.33\columnwidth]{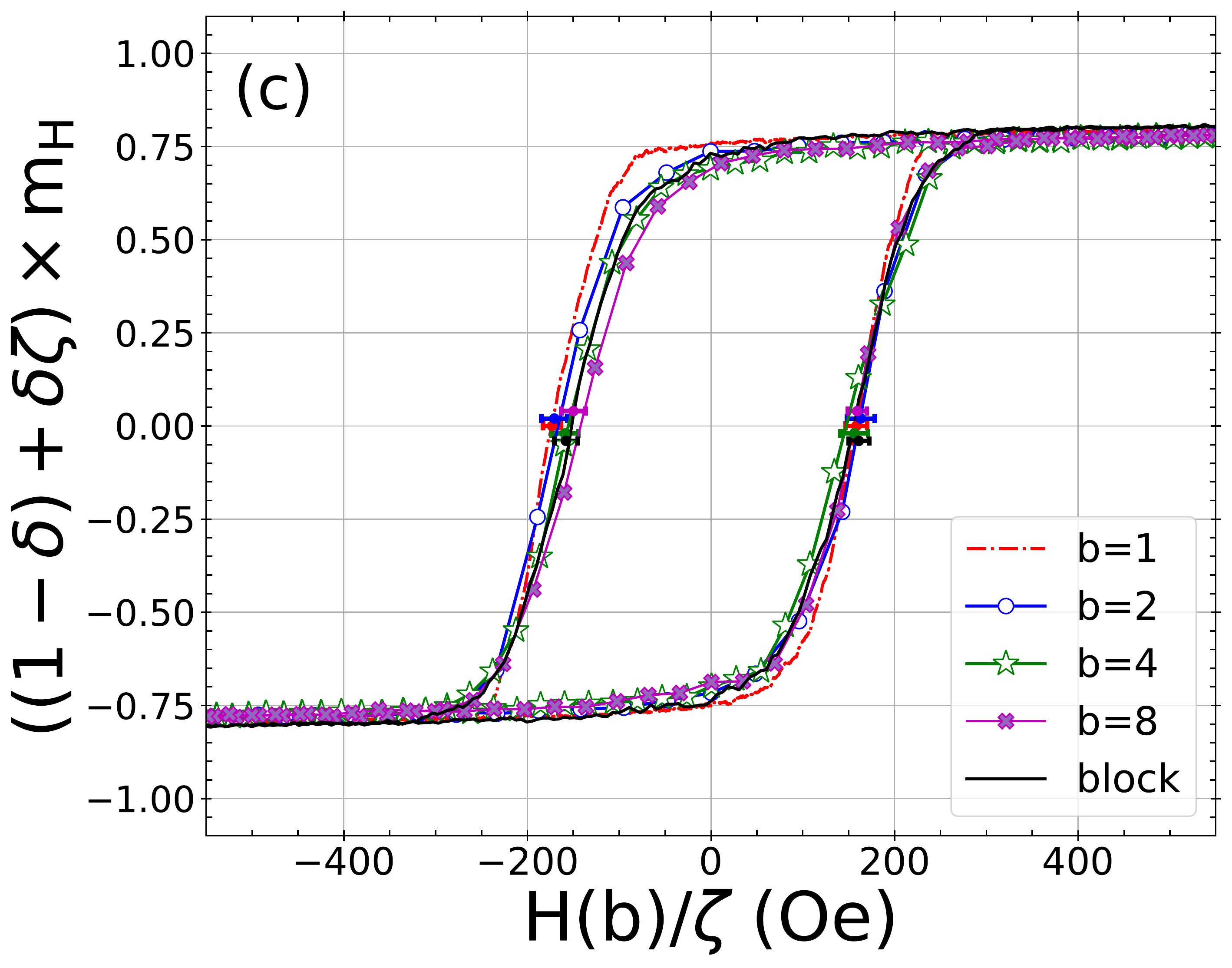}
  \caption{Application of RG coarse graining to nanorod MH loops at $T=310$~K and SR$=2.5$~Oe/ns. (a) Changing cell length ($a=ba_0$) without changing magnetic parameters. (b) $A$ and $K$ are scaled according to Eqs.~\ref{eq:RNG_scalingA} and~\ref{eq:RNG_scalingK}, respectively, and $m_H$ and $H$ are scaled according to Eqs.~\ref{eq:RNG_scalingM} and~\ref{eq:RNG_scalingH}, respectively. (c) As in panel (b), except $m_H$ is scaled according to Eq.~\ref{eq:corrM0} with $\delta$=0.511.  
  $\Delta t=1$~fs for all simulations. 
  Horizontal error bars shown for $H_c$ represent one standard error and are vertically displaced to avoid overlap.
  Uncertainty in $H_c$ is approximately 7 to 13\%.
  }\label{fig:RNG_1}
\end{figure*}

In Fig.~\ref{fig:RNG_1}a we plot hysteresis loops at $T=310$~K using different cell sizes (varying $b$) while keeping the exchange and anisotropy parameters fixed at $A_0$ and $K_0$.  A value of ${\rm SR}=2.5$~Oe/ns is chosen to make the simulations computationally feasible at $b=1$.  
Both the coercivity $H_c$ and the remanence increase with increasing $b$.  The increasing loop area is consistent with the stronger exchange coupling ($J_{\rm eff}= b a_0 A_0$) between magnetization vectors of adjacent cells.  For $b \ge 4$, it appears that the exchange is strong enough for the system to be nearly uniformly magnetized, and so  $H_c$ remains largely unchanged for $b\ge 4$ since $K$ is constant.  This means that for $b=1$, at this $T$ and for our rod size, exchange is not strong enough to be able to treat the nanorod as a single macrospin in a trivial way.  Clearly, varying cell size changes the loops and a coarse-graining procedure is required.



In their coarse-graining procedure,
Grinstein and Koch introduced a reduced temperature $T^*$, which for a three dimensional system is given by,
\begin{equation}\label{eq:reducedT}
    T^*=\frac{k_B T \Lambda}{A}.
\end{equation}
where $\Lambda=2\pi/ba_0$ is a high wave-number cut-off that reflects the level of coarse-graining. 
Similarly, the reduced parameters for field and anisotropy constants are defined as,
\begin{equation}
        h=\frac{\mu_0 M_s H}{A\Lambda^2} \frac{1000}{4\pi}, \qquad  g=\frac{K}{A\Lambda^2},
\end{equation}
with $H$ given in Oe.
Introducing the parameter $l=\ln(b)$, they gave the following set of equations for calculating the reduced parameters as functions of cell size,
\begin{equation}\label{eq:diffVariations}
\eqalign{
         \frac{dT^*(l)}{dl}=\left[-1+F  \left(T^*(l),h(l),g(l)\right)\right] T^*(l)\cr 
         \frac{dh(l)}{dl}=2h(l)  \cr
         \frac{dg(l)}{dl}=\left[2-2F\left(T^*(l),h(l),g(l)\right)\right]g(l)}
\end{equation}
where
\begin{equation}
    F(T^*, h, g)=\frac{T^*}{2\pi(1+h+g)}.
\end{equation}
Additionally, the magnetization of the coarse-grained system is scaled via,
\begin{equation}
    M(T^*,h)=\zeta(l)\times M(T^*(l),h(l))
\end{equation}
where 
\begin{equation}
    \zeta(l)=e^{-\int_0^{l}F(T^*(l^{\prime}),h(l^{\prime}),g(l^{\prime}))dl^{\prime}}.
\end{equation}

For our system parameters and range of $H$, both $g\ll1$ and $h\ll 1$, and so $F\simeq T/2\pi$, which makes the numerical solution of Eq.~\ref{eq:diffVariations} practically indistinguishable from the approximate analytic solution, which we find to be,
\begin{eqnarray}
    A(b) &=& \zeta(b) \times A_0 \label{eq:RNG_scalingA}\\
    K(b) &=& \zeta(b)^3 \times K_0 \label{eq:RNG_scalingK}\\
    H(b) &=& \zeta(b) \times H_0 \label{eq:RNG_scalingH}\\
    M_0 &=& \zeta(b) \times M(b) \label{eq:RNG_scalingM}
\end{eqnarray}
where
$t=T/T_c$ and $\zeta(b)=t/b+1-t$.
At $T=310$~K, $t=0.3613$, $\zeta(2)=0.8193$,
$\zeta(4)=0.7290$, $\zeta(8)=0.6839$, and $\zeta(22)=0.6551$.

Eqs.~\ref{eq:RNG_scalingA} and~\ref{eq:RNG_scalingK} provide a prescription for changing material parameters with $b$, while Eqs.~\ref{eq:RNG_scalingH} and~\ref{eq:RNG_scalingM} provide the prescription for scaling $H$ and $M$ after a loop calculation.  However, we find that the prescription does not yield loops that are invariant with $b$, on account of Eq.~\ref{eq:RNG_scalingM}; the correction of the coarse-grained values of $M$ back to those corresponding to the unscaled system is too large (the corrected remanance is too small), as we show in Fig.~\ref{fig:RNG_1}b.
In Fig.~\ref{fig:RNG_1}c, we apply a correction to Eq.~\ref{eq:RNG_scalingM} and obtain good agreement between the reference ($b=1$) and coarse-grained ($b>1$) loops.

To motivate our correction to the  rescaling of the magnetization, we begin by noting that the same value of $T^*$ in Eq.~\ref{eq:reducedT} can be achieved by either having a rescaled temperature $T(b)$ or having a  rescaled $A(b)$.  Combining this idea with Eq.~\ref{eq:RNG_scalingA} yields,
\begin{equation}\label{eq:reducedT_b}
    T(b)= \frac{T_0}{b\zeta(b, T_0)},
\end{equation}
which together with Eq.~\ref{eq:RNG_scalingM} [after solving for $M(b)$] predicts an overly simple dependence of $M$ on $T$, parametrically through $b$: a line passing through $M_0$ and $T_0$ at $b=1$ and through $M=0$ and $T=T_c$ as $b\rightarrow 0$.

\begin{figure}\centering
\includegraphics[width=0.8\columnwidth]{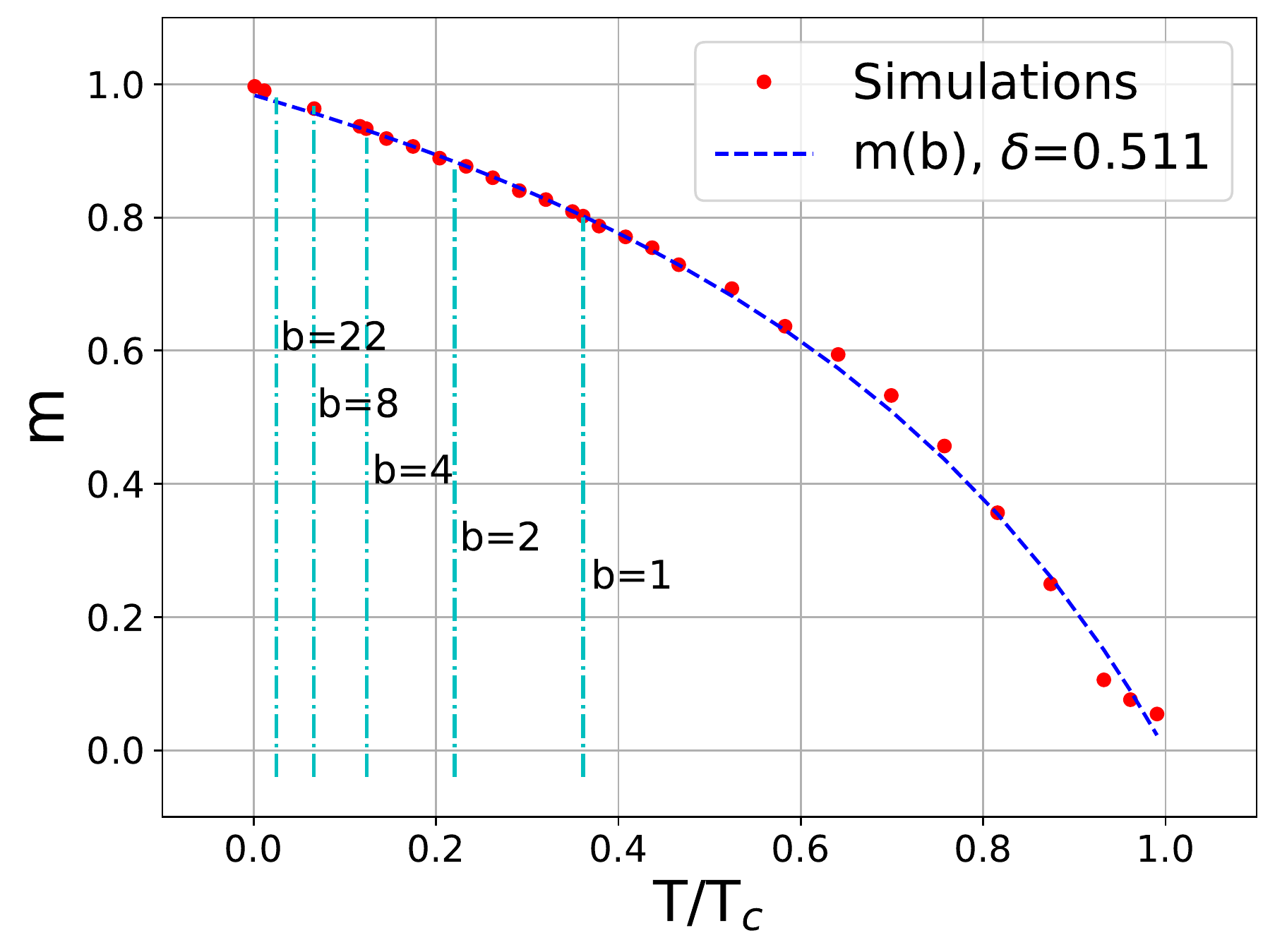}
\caption{
Determining a scaling function for $M(b)$ from the $T$ dependence of the nanorod magnetization.
$\delta$ is used as a fitting parameter to match nanorod data, yielding a value of 0.511.
Vertical dot-dash lines indicate reduced temperatures corresponding to different values of $b$.}
    \label{fig:M_T}
\end{figure}

To obtain a model that better matches the data, we introduce a phenomenolgical correction to Eq.~\ref{eq:RNG_scalingM}, one in which $M_0$ is a weighted average of $M(b)$ and the RG expression for $M_0$,
\begin{equation}\label{eq:corrM0}
M_0 = \delta \zeta(b, T_0) M(b) + (1-\delta) M(b).
\end{equation}
We use $\delta$ as a free parameter to fit the $M(T)$ data for the nanorod. 
This yields a value of $\delta=0.511$, which we use in rescaling $m_H$ in Fig.~\ref{fig:RNG_1}c. The fit reasonably recovers $M(T)$ in the $T$ range corresponding to values of $b$ between 1 and 22, as shown in Fig.~\ref{fig:M_T}.  

The collapse of the data in Fig.~\ref{fig:RNG_1}c is remarkable, with the biggest discrepancy arising between $b=1$, corresponding to the most fine-grained simulation, and $b=2$, the first step in coarse-graining.  The difference lies most noticeably in the shoulder region where  magnetization begins to change, where the microscopic details likely matter most.  Loss of some detail is expected with coarse-graining and consistent with previous studies involving atomic-level magnetization switching in a grain~\cite{mercer2011atomic}. 
The magnetization in the shoulder areas appears to diminish with increasing $b$.  
{\red The behavior of $b=22$ runs counter to this trend, but at this level of coarse-graining, there is only a single cell. 
It is significant, however, that scaling seems to hold even in this limit.
(We note that in this limit, even though there are no exchange interactions in the simulations, the value of the effective anisotropy still depends on exchange through the dependence of $T_c$ on $A_0$.)}
The loop areas for $b=1$, 2 , 4, 8 and 22 are 
495, 488, 443, 432 and 472 Oe, 
respectively. The smallest loop area (for $b=8$) is 13\% smaller than the area for $b=1$.

%

{\red We note that the unrenormalized exchange length for our simulated material is $l_{\mathrm {ex,0}}=\sqrt{\frac{2A_0}{\mu_0 M_s^2}}=8.23$~nm, which is longer than $a_8=6.712$~nm, and so only our $b=22$ single block simulations scale the cell size beyond $l_{\rm ex,0}$.  Under renormalization, however, the exchange length becomes 
$l_{\rm ex,b} = \sqrt{\frac{2 \zeta(b) A_0}{\mu_0M_s^2}}$, which
decreases with increasing $b$, and takes on values
7.45, 7.02, 6.80 and 6.66~nm for $b=2, 4, 8,$ and 22, respectively.  Thus for $b=8$, the cell length and the exchange length are approximately the same.}

We now turn our attention to speeding up simulations by considering the relationship between SR and  $\alpha$.
A larger value of $\alpha$ signifies a faster loss of energy and a shorter relaxation time for alignment of the magnetic moments to the field, and results in a smaller hysteresis loop. Likewise, a slower SR is equivalent to a longer measurement time 
and consequently a smaller hysteresis loop. 
To build on these ideas, we recall Sharrock's equation for $H_c$ as a function of $T$~\cite{Sharrock_1981},
\begin{equation}\label{eq:Sharrock}
    H_c=H_K\left[ 1-\sqrt{\frac{k_BT}{KV}\ln\left(\frac{f_0\tau}{\ln 2}\right)}\,\,\right].
\end{equation}
Sharrock derived this equation by calculating the time required for half of the magnetization vectors in the system, which are initially anti-aligned with the field, to overcome an energy barrier that grows with $KV$ and align with a field of strength $H_c$.  In this context, $\tau$ is the relaxation time.  In the context of hysteresis loops, $H_c$ is the field required to flip half of the magentization vectors in an observation time $\tau$, which is related to SR via $\tau \propto 1/{\rm SR}$.  $f_0$ 
is the so-called attempt frequency, for which   Brown~\cite{brown1963thermal, garcia1998langevin, breth2012thermal, taniguchi2012thermal, leliaert2015vinamax} derived an expression in the high-barrier limit.
At small $\alpha$, $f_0\propto \alpha$, and so the product $f_0 \tau \propto \alpha/{\rm SR}$, implying that so long as ${\rm SR}/\alpha = \mathrm{constant}$, $H_c$ should remain the same.

\begin{figure}
\centering
\includegraphics[width =0.8\columnwidth]{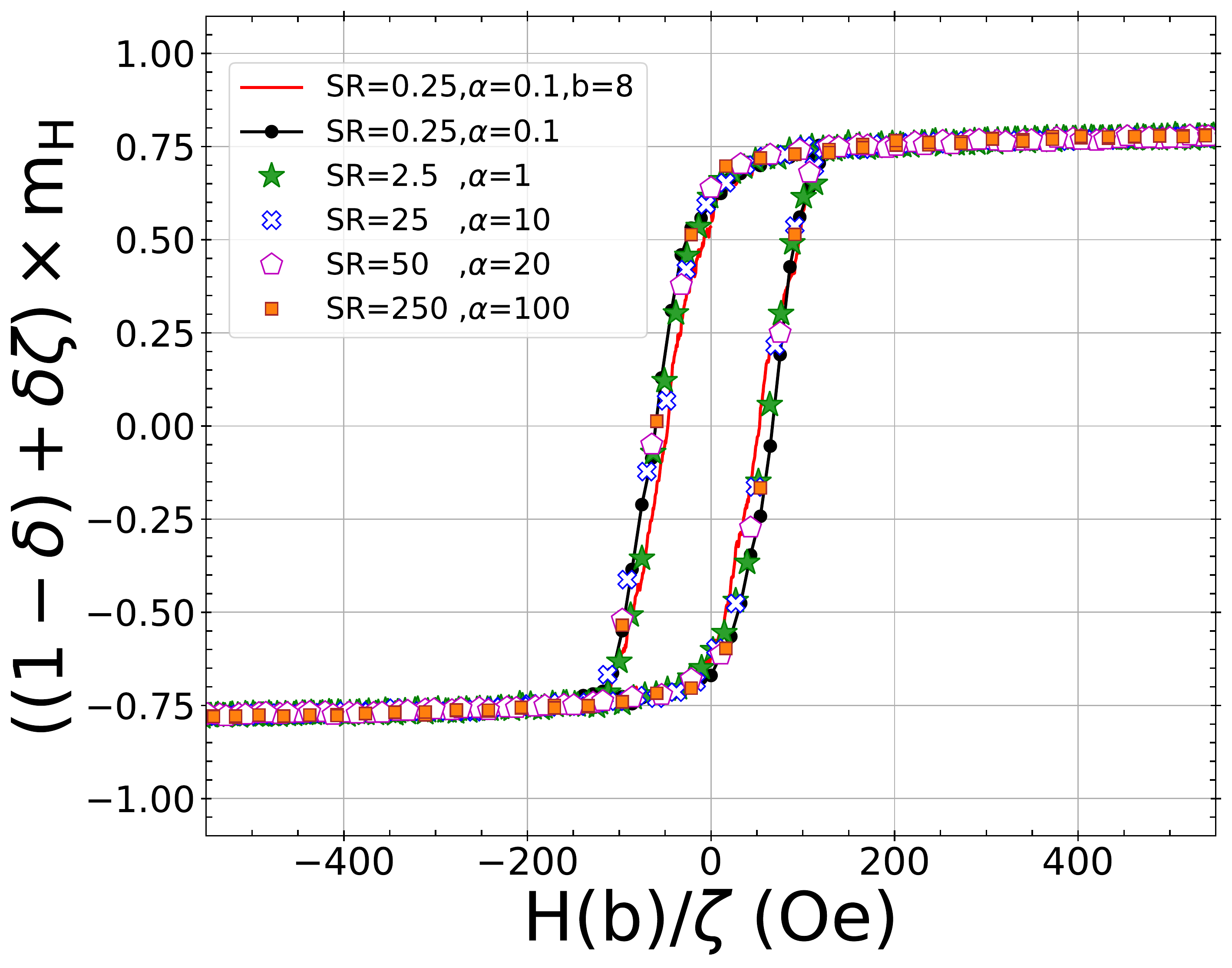}
\caption{Invariance of MH loops. We combine RG scaling of magnetic quantities, larger time step with block size, and ${\rm SR}/\alpha$ scaling to predict the behaviour of prohibitively long  fine-grain ($b=1$) simulations. $b=4$ unless otherwise noted.
}\label{fig:SR_Alpha}
\end{figure}

In Fig.~\ref{fig:SR_Alpha} we show loops calculated for ${\rm SR}/\alpha=2.5$ ($H_{\rm max}=500$~Oe, and $f=125$~kHz), the ratio obtained using a clinically relevant ${\rm SR}=0.25$~Oe/ns and the estimate of $\alpha=0.1$.
Data for $b=4$ and 8 and for various SR-$\alpha$ pairs show good agreement.   At $0.25$~Oe/ns, simulations using $b=1$ are prohibitively long, taking several months on available computing resources.  The results shown here combine the RG approach to reduce the number of cells, the ability to use a larger time step $\Delta t$ for larger cells in solving the LLG equation~\cite{Wang20112676}, and the ${\rm SR}/\alpha$ scaling to employ a faster SR, all to dramatically reduce simulation time --  by a factor of $4^3$ to $8^3$ for reducing the number of cells, a factor of at least 5 for the time step, and a factor of up to 1000 when using the fastest SR.
The average area of the five loops for $b=4$ in {\red Fig.~\ref{fig:SR_Alpha}} is $S=171.3 \pm 2.8$~Oe, translating to a specific loss power of $f \mu_0 \frac{1000}{4\pi} M_s S/ \rho  = 207$~W/g $\pm 10\%$ (using $\rho=5.17$~g/cm$^3$), which is consistent with clinical expectations~\cite{das2019recent}.  The loop area for $b=8$ is 13\% lower at $149.4$~Oe.



In summary, we show that our modification to the RG approach of Grinstein and Koch~\cite{grinstein2003coarse} yields a scaling of exchange and anisotropy parameters and finite temperature nanorod hysteresis loops that are, to  approximately 10-15\%, invariant with cell size.  
{\red We note that the coarse-graining of magnetostatic interactions is beyond the framework of Ref.~\cite{grinstein2003coarse}.
We are currently investigating magnetostatic scaling, and intend to report on it in future work.}

Scaling results hold even to the point where the nanorod is represented by a single magnetization vector 
that experiences anisotropy only.  Whether this limit holds for systems with weaker exchange remains to be studied.
This reduction to an effective  Stoner-Wohlfarth (SW) model~\cite{stoner1948mechanism} should facilitate comparison with experiments on nanorods, since an analytic solution to the SW model at finite $T$ and SR exists~\cite{usov2010low}. It should also simplify computational studies of nanoparticles (nanorod composites) and collections of nanoparticles
used in a wide variety of applications and hence facilitate comparison with experimental MH loops and quantification of system properties through simulations.

In addition to the computational speedup resulting from the use of fewer micromagnetic cells, the invariance of loops when ${\rm SR}/\alpha$ is fixed provides another avenue for computational speedup by allowing one to use a larger SR than the target value.  We caution, however, that the theoretical motivation for this invariance stems from considering the Sharrock equation 
for only small $\alpha$.
While both SR and $\alpha$ set time scales, we have not provided any  reasoning for why the invariance should hold as well as it does for larger $\alpha$. 

The data that support the findings of this study are available from the corresponding author upon reasonable request.


\ack{
We thank Johan van Lierop, Rachel Nickel and Mikko Karttunen for enlightening discussions, and Martin D.~Leblanc for guidance in using OOMMF.  R.B. and I.S.-V. thank Mikko Karttunen and Styliani Consta for hosting our stay at Western University.  We acknowledge the financial support from the Natural Sciences and Engineering Research Council (Canada).  Computational resources were provided by ACENET and Compute Canada.}

\section*{References}

\bibliographystyle{iopart-num}
\bibliography{ref-num}

\providecommand{\newblock}{}
\begin{thebibliography}{10}
\expandafter\ifx\csname url\endcsname\relax
  \def\url#1{{\tt #1}}\fi
\expandafter\ifx\csname urlprefix\endcsname\relax\def\urlprefix{URL }\fi
\providecommand{\eprint}[2][]{\url{#2}}

\bibitem{abo2013definition}
Abo G~S, Hong Y~K, Park J, Lee J, Lee W and Choi B~C 2013 {\em IEEE Trans.
  Magn.\/} {\bf 49} 4937--4939

\bibitem{kirschner2006relaxation}
Kirschner M, Schrefl T, Hrkac G, Dorfbauer F, Suess D and Fidler J 2006 {\em
  Physica B\/} {\bf 372} 277--281

\bibitem{kirschner2005cell}
Kirschner M, Schrefl T, Dorfbauer F, Hrkac G, Suess D and Fidler J 2005 {\em
  J.~Appl.~Phys.\/} {\bf 97} 10E301

\bibitem{CoarseGrainingFengVisscher}
Feng X and Visscher P~B 2001 {\em J.~Appl.~Phys.\/} {\bf 89} 6988--6990

\bibitem{gilbert2004phenomenological}
Gilbert T~L 2004 {\em IEEE Trans. Magn.\/} {\bf 40} 3443--3449

\bibitem{Wang20112676}
Wang X, Gao K and Seigler M 2011 {\em IEEE Transactions on Magnetics\/} {\bf
  47} 2676--2679

\bibitem{grinstein2003coarse}
Grinstein G and Koch R~H 2003 {\em Phys. Rev. Lett.\/} {\bf 90} 207201

\bibitem{Westmoreland2018266}
Westmoreland S, Evans R, Hrkac G, Schrefl T, Zimanyi G, Winklhofer M, Sakuma N,
  Yano M, Kato A, Shoji T, Manabe A, Ito M and Chantrell R 2018 {\em Scripta
  Materialia\/} {\bf 154} 266--272

\bibitem{dennis2009nearly}
Dennis C, Jackson A, Borchers J, Hoopes P, Strawbridge R, Foreman A, Van~Lierop
  J, Gr{\"u}ttner C and Ivkov R 2009 {\em Nanotechnology\/} {\bf 20} 395103

\bibitem{victora2003effects}
Victora R, Willoughby S, MacLaren J and Xue J 2003 {\em IEEE Trans. Magn.\/}
  {\bf 39} 710--715

\bibitem{heider1988note}
Heider F and Williams W 1988 {\em Geophys.~Res.~Lett.\/} {\bf 15} 184--187

\bibitem{kouvel1956specific}
Kouvel J 1956 {\em Phys.~Rev.\/} {\bf 102} 1489

\bibitem{moskowitz1987theoretical}
Moskowitz B~M and Halgedahl S~L 1987 {\em J.~Geophys.~Res.~Solid Earth\/} {\bf
  92} 10667--10682

\bibitem{glasser1963spin}
Glasser M~L and Milford F~J 1963 {\em Phys.~Rev.\/} {\bf 130} 1783

\bibitem{srivastava1979exchange}
Srivastava C~M, Srinivasan G and Nanadikar N~G 1979 {\em Phys. Rev. B\/} {\bf
  19} 499

\bibitem{srivastava1987spin}
Srivastava C and Aiyar R 1987 {\em J.~Phys.~C: Solid St.~Phys.\/} {\bf 20} 1119

\bibitem{uhl1995first}
Uhl M and Siberchicot B 1995 {\em J. Phys. Condens. Matter\/} {\bf 7} 4227

\bibitem{dutz2013magnetic}
Dutz S and Hergt R 2013 {\em Int.~J.~Hyperth.\/} {\bf 29} 790--800

\bibitem{usov2013properties}
Usov N, Gudoshnikov S, Serebryakova O, Fdez-Gubieda M, Muela A and
  Barandiar{\'a}n J 2013 {\em J.~Supercond.~Nov.~Magn.\/} {\bf 26} 1079--1083

\bibitem{usov2017interaction}
Usov N, Serebryakova O and Tarasov V 2017 {\em Nanoscale research letters\/}
  {\bf 12} 1--8

\bibitem{plumer2010micromagnetic}
Plumer M, van Lierop J, Southern B and Whitehead J 2010 {\em J. Phys. Condens.
  Matter\/} {\bf 22} 296007

\bibitem{cullity2011introduction}
Cullity B~D and Graham C~D 2011 {\em Introduction to magnetic materials\/}
  (John Wiley \& Sons)

\bibitem{Newell1993}
Newell A~J, Williams W and Dunlop D~J 1993 {\em Journal of Geophysical
  Research: Solid Earth\/} {\bf 98} 9551--9555

\bibitem{Aharoni1998}
Aharoni A 1998 {\em Journal of Applied Physics\/} {\bf 83} 3432--3434

\bibitem{fukushima1998volume}
Fukushima H, Nakatani Y and Hayashi N 1998 {\em IEEE Trans. Magn.\/} {\bf 34}
  193--198

\bibitem{usov2012magnetic}
Usov N and Barandiar{\'a}n J 2012 {\em Journal of Applied Physics\/} {\bf 112}
  053915

\bibitem{usov2010low}
Usov N 2010 {\em J.~Appl.~Phys.\/} {\bf 107} 123909

\bibitem{serrano2011thickness}
Serrano-Guisan S, Wu H~C, Boothman C, Abid M, Chun B, Shvets I and Schumacher H
  2011 {\em J.~Appl.~Phys.\/} {\bf 109} 013907

\bibitem{hergt2007magnetic}
Hergt R and Dutz S 2007 {\em J. Magn. Magn. Mater.\/} {\bf 311} 187--192

\bibitem{invspinel}
 2019 {\em Introduction to Inorganic Chemistry\/} (Wikibooks) chap 8.6, see
  Creative Commons licence http://creativecommons.org/licenses/by-nc-sa/3.0/us/
  \urlprefix\url{https://en.wikibooks.org/wiki/Introduction_to_Inorganic_Chemistry}

\bibitem{brown1963thermal}
Brown~Jr W~F 1963 {\em Phys.~Rev.\/} {\bf 130} 1677

\bibitem{OOMMF}
Donahue M~J and Porter D~G 1999 {\em OOMMF User's Guide, Version 1.0,
  Interagency Report NISTIR 6376\/} National Institute of Standards and
  Technology Gaithersburg, MD \urlprefix\url{https://math.nist.gov/oommf/}

\bibitem{theta_evolve}
Lemcke O 2004 {\em ThetaEvolve for OOMMF releases: 1.2a3,\/} see
  https://math.nist.gov/oommf/contrib/oxsext/oxsext.html
  \urlprefix\url{http://www.nanoscience.de/group_r/stm-spstm/projects/temperature/download.shtml}

\bibitem{mercer2011atomic}
Mercer J, Plumer M, Whitehead J and Van~Ek J 2011 {\em Appl.~Phys.~Lett.\/}
  {\bf 98} 192508

\bibitem{Sharrock_1981}
{Sharrock} M and {McKinney} J 1981 {\em IEEE Trans. Magn.\/} {\bf 17}
  3020--3022 ISSN 0018-9464

\bibitem{garcia1998langevin}
Garc{\'\i}a-Palacios J~L and L{\'a}zaro F~J 1998 {\em Phys. Rev. B\/} {\bf 58}
  14937

\bibitem{breth2012thermal}
Breth L, Suess D, Vogler C, Bergmair B, Fuger M, Heer R and Brueckl H 2012 {\em
  J.~Appl.~Phys.\/} {\bf 112} 023903

\bibitem{taniguchi2012thermal}
Taniguchi T and Imamura H 2012 {\em Phys. Rev. B\/} {\bf 85} 184403

\bibitem{leliaert2015vinamax}
Leliaert J, Vansteenkiste A, Coene A, Dupr{\'e} L and Van~Waeyenberge B 2015
  {\em Med.~Biol.~Eng.~Comput.\/} {\bf 53} 309--317

\bibitem{das2019recent}
Das P, Colombo M and Prosperi D 2019 {\em Colloids and Surfaces B:
  Biointerfaces\/} {\bf 174} 42--55

\bibitem{stoner1948mechanism}
Stoner E~C and Wohlfarth E 1948 {\em Philos.~Trans.~Royal Soc.~A\/} {\bf 240}
  599--642

\end{thebibliography}

\end{document}